# Progress in lattice chiral gauge theories

R. Narayanan[a] [*] and H. Neuberger[b]

[a]University of Washington, Institute for Nuclear Theory, Box 351550, Seattle, WA 98195-1550

[b]Dept. of Physics and Astronomy, Rutgers University, Piscataway, NJ 08855-0849

Some key features of continuum chiral fermions are shown to be satisfied by the overlap.

The central character in the chiral fermion action is the chiral differential operator operating on Weyl fermions, $\mathbf{C}(A)$. The fermion is in some irreducible representation of the gauge field $A$. It is best to avoid infrared complications and work on a compact manifold like a torus. The operator $\mathbf{C}(A)$ has some special properties, established in the continuum [1].

**1.** *No eigenvalue problem:* $\mathbf{C}(A)$ is a map between two spaces carrying different representations of the Euclidean Lorenz group. When $A$ is replaced by $A'$ as a result of a Lorenz transformation, $\mathbf{C}(A)$ does not transform by a similarity transformation. One cannot associate a Lorenz invariant eigenvalue problem with the linear operator $\mathbf{C}(A)$. (The associated Dirac operator, $\mathbf{D}(A)$, in the Weyl representation is made out of $\mathbf{C}(A)$ and $\mathbf{C}^\dagger(A)$ in the off-diagonal sub-blocks and has a Lorenz invariant eigenvalue problem.)

**2.** *Topology of gauge fields:* The space of the gauge fields splits into disconnected blobs labeled by an integer $q$, the topological charge. $\mathbf{C}(A)$ acts on functions when $q = 0$, but for $q \ne 0$ $\mathbf{C}(A)$ acts on slightly more general objects, which are, roughly, only functions up to some special gauge transformations. When $q = 0$ one can think about $\mathbf{C}(A)$ as an infinite square matrix, but when $q \ne 0$ the appropriate view is that $\mathbf{C}(A)$ is an infinite rectangular matrix and $q$ determines the difference between the number of rows and the number of columns. Therefore the result of integration of the fermionic action over the fermion fields, viz., the determinant of $\mathbf{C}(A)$, is non-zero only in the $q = 0$ blob.

**3.** *Anomalies:* There is usually an obstruction to making the determinant of $\mathbf{C}(A)$ gauge invariant over the $q = 0$ blob. The obstruction can be understood by considering certain two dimensional disks in the $A$-space whose boundaries lie in their entirety on single gauge orbits. These disks have an interior point (or points) where $\det \mathbf{C}(A)$ vanishes. Otherwise, $\det \mathbf{C}(A)$ behaves generically and its phase therefore winds around the boundary of the disk. This implies that one cannot choose a constant phase on the boundary of the disk and hence the obstruction. In such cases the theory is said to be anomalous. Anomalies can be cancelled between different irreducible representations and one can construct an anomaly free fermion representation by a suitable combination of irreducible representations. Vector gauge theories are special cases of anomaly free chiral gauge theories.

When we regularize a chiral gauge theory, we start by accepting the formal factorization of the Grassmann integral into contributions per individual irreducible multiplet in the continuum as an exact property. Next we wish to preserve the formal continuum properties listed above.

Property 1 indicates that it is not advisable to write the determinant of $\mathbf{C}(A)$ as a product of eigenvalues and then regulate that product. Property 2 implies that $\mathbf{C}(A)$ is not even a square matrix. On a lattice, and likely in any other completely defined regularization, any non-trivial topology present in the continuum will be lost because its roots are always in the smoothness of the underlying space–time torus. But there is a remnant of topology on the lattice and one can associate a topological charge with gauge fields on a finite lattice in a manner that is consistent with the continuum definition. Therefore the physical effects of topology should be reproduced on the lattice. For example, in a vector gauge theory with $N_f$ massless fermions, the $U_R(N_f) \times U_L(N_f)$ global symmetry formally present in the massless case gets broken down to $SU_R(N_f) \times SU_L(N_f) \times U_V(1)$[2]. This explicit breaking (as opposed to spontaneous) occurs because certain Fermi operators, consisting of $N_f$ left handed fields times $N_f$ right handed fields, known as 't Hooft vertices, acquire a non–zero expectation values in gauge backgrounds carrying unit topological charge. In anomaly free chiral theories (as opposed to vector theories) the 't Hooft vertices can induce fermion number violating processes.

---
[*]Speaker

Nonzero topological charges are dynamically produced in Monte Carlo simulations and this is consistent with the fact that lattice theories should obey clustering. In the continuum, clustering ties zero topology sectors to the other sectors ensuring that all sectors contribute. Thus, one must get right the physics for nonzero lattice topology. The rectangular nature of $\mathbf{C}(A)$ is essential to make this work. If a regularization scheme forces $\mathbf{C}(A)$ to be a square matrix for all gauge fields, such a scheme ends up packing several rectangular $\mathbf{C}(A)$, $\mathbf{C}^\dagger(A)$ structures into the rigidly square shaped allotted space when the gauge field has $q \neq 0$. So, "doubling" can be viewed as coming from the $q \neq 0$ blobs! Since in any regularization scheme the different $q$ blobs are connected by thin "necks" the doubling seen for $q \neq 0$ fields can also be seen in the $q = 0$ sector. Therefore any regularization scheme where one has a candidate square matrix representing $\mathbf{C}(A)$ in a fixed template is bound to go wrong.

Let us be very explicit and consider the $N_f = 1$ vector theory. $<\bar{\psi}\psi> \neq 0$ is explained by diagonalizing the Dirac operator $D$ and observing that it has a zero eigenmode in a background $A$ carrying unit topological charge. $<\bar{\psi}\psi>_A$ gets a non vanishing contribution from $\int d\bar{\psi}_n d\psi_n e^{-\sum_n \lambda_n \bar{\psi}_n \psi_n} \bar{\psi}_0 \psi_0$ where $\lambda_0 = 0$. When the integral is factorized into the right and left pieces we see that we need $<\bar{\psi}_R>_A$ and $<\psi_L>_A$ nonzero (or the same after exchanging $L$ and $R$). This can happen only when there is a deficiency of $\bar{\psi}_R$ integration variables relative to $\psi_R$ and of $\psi_L$ relative to $\bar{\psi}_L$. This deficiency is most usefully viewed as a result of the chiral blocks of $D$ being rectangular, packed together in a square, diagonalizable structure. This type of "packaging" is typical of the vector theory.

In anomaly free theories, property 3 implies that $\mathbf{C}(A)$ can be made gauge invariant. This will be the case when the regulator for the fermions is removed keeping the smooth background gauge field fixed. On the lattice one has a common regulator (the lattice spacing) for all fields and the continuum limit is taken by tuning the coupling constants. Typically, the determinant of $\mathbf{C}(A)$ will not be gauge invariant as long as the lattice spacing is finite. However, on a finite lattice the group of all local gauge transformations is truly compact and gauge invariance is always trivially achievable by group averaging. So, as shown in detail by Foerster, Nielsen and Ninomiya [3], the issue of having an exactly gauge invariant action is largely a matter of taste rather than substance. The dynamically relevant question is whether, in "typical" backgrounds, the gauge transformation degrees of freedom are short range correlated or not. In anomaly free cases they should be, while the Wess–Zumino action indicates that in the anomalous case they are not. Whether in the latter case a larger continuum theory encompassing them exists or not is an open question. As emphasized by Faddeev and Shatashvili [4], at the classical level there is a nice phase space structure; unfortunately, we do not know how to quantize the theory beyond the formal level. We believe that any credible approach of the "factorized" type to the problem of non-perturbatively defining chiral gauge theories will provide a tool applicable also to the Faddeev–Shatashvili problem.

The overlap formalism [5] for the construction of lattice chiral gauge theories provides a formula for the computation of the chiral determinant and any fermionic n-point function in a fixed gauge background. The formula is written down with the formal properties of $\mathbf{C}(A)$ in mind. It is first written down at a formal level and it is subsequently regularized on the lattice. Very nice work on the overlap has been done by Randjbar–Daemi, Strathdee and Fosco in [6].

The overlap formula is to be inserted in an inte-gral over all gauge fields. The formula is provided for a single chiral fermion with a definite chirality. Theories are constructed by putting together many such blocks. Let $U$ denote a fixed gauge background on a finite lattice. Two single particle Hamiltonians

$$\mathbf{H}^\pm(U) = \begin{pmatrix} \mathbf{B}^\pm(U) & \mathbf{C}(U) \\ \mathbf{C}^\dagger(U) & -\mathbf{B}^\pm(U) \end{pmatrix}$$

$$\mathbf{C}(U; x\alpha i, y\beta j) = \frac{1}{2}\sum_\mu \sigma_\mu(\alpha, \beta)$$
$$\left[\delta_{y,x+\hat{\mu}} U_\mu^{ij}(x) - \delta_{x,y+\hat{\mu}} [U_\mu^\dagger(y)]^{ij}\right]$$

$$\mathbf{B}^\pm(U; x\alpha j, y\beta k) = \frac{1}{2}\delta_{\alpha,\beta}\left\{\sum_\mu \left[2\delta_{x,y}\delta_{ij} - \delta_{y,x+\hat{\mu}} U_\mu^{ij}(x) - \delta_{x,y+\hat{\mu}} [U_\mu^\dagger(y)]^{ij}\right] \pm m\delta_{ij}\delta_{xy}\right\}$$

are associated with the background gauge field. $\mathbf{C}(U)$ is the lattice version of the chiral Dirac operator and the above Hamiltonians are defined to provide a definition for the determinant of this operator. Details regarding the notation in the above equation can be found in section 7 of [5]. The matrices $\mathbf{H}^\pm(U)$ are square with the linear size equal to $4 \times L^4 \times N$ in a 4-D $L^4$ lattice and the fermion is assumed to be in the fundamental representation of the gauge group, $SU(N)$.

$$\mathcal{H}^\pm(U) = \sum_{x\alpha i, y\beta j} a^\dagger_{x\alpha i} \mathbf{H}^\pm(x\alpha i, y\beta j; U) a_{y\beta j}$$

are two many body Hamiltonians for non-interacting fermions with the single particle Hamiltonians given by $\mathbf{H}^\pm(U)$. $a^\dagger_{x\alpha i}$ and $a_{y\beta j}$ are fermion creation and destruction operators that obey the usual canonical anti-commutation relations. With these definitions in place

$$\det \mathbf{C}(U) \Leftrightarrow {}^{WB}_U\!< L - | L+ >^{WB}_U$$
$$\det \mathbf{C}^\dagger(U) \Leftrightarrow {}^{WB}_U\!< R - | R+ >^{WB}_U$$

are appropriate definitions for the chiral determinants. $|L\pm >^{WB}_U$ are the many body ground states of $\mathcal{H}^\pm(U)$ and $|R\pm >^{WB}_U$ are the many body ground states of $-\mathcal{H}^\pm(U)$. Fermionic n-point functions are defined by the appropriate insertion of fermion creation and destruction operators inside the overlap of the two many body states. Details concerning the n-point functions can be found in section 5 of [5]. Since the formula is an overlap of two different many body states it is necessary to fix the phase of these states to completely define the chiral determinant. The phases are fixed according to the Wigner-Brillouin choice, hence the superscript $WB$ on the phases. In the trivial topological sector, this choice means that ${}_1\!< L\pm | L\pm >^{WB}_U$ are real for all $U$ in this sector. Extension of this definition to all sectors can be found in section 5.2 of [5]. Note that the chiral determinant is defined as a complex functional of the gauge background $U$. One can show that this definition is not gauge invariant and that a gauge variation of the background gauge field produces the *consistent* anomaly. This is shown numerically in 2-D in section 10.2 of [5] and section 11.1 shows the same in 4-D. Analytically this is established in [6]. Many other properties expected of the chiral determinant and the fermionic n-point functions are obeyed by the overlap formula. Section 4 in [5] shows that the formal properties of the phase of the chiral determinant are obeyed by the overlap formula. Section 10 and 11 in [5] illustrate many tests done on the overlap formula.

A key feature of the overlap formalism is the mixed nature present in the definition: Fermions are treated in an auxiliary operator language and gauge fields appear in the usual path-integral form. This is not unusual, after all the Grassmann integrals used to incorporate fermions in the path integral are carried out analytically before numerical simulation are attempted. Similarly here only the final formulae are accessible to numerical simulation. One diagonalizes the single particle Hamiltonians, $\mathbf{H}^\pm(U)$. The many body ground states consist of all the eigenvectors of $\mathbf{H}^\pm(U)$ with negative eigenvalues. An overlap matrix is formed by computing the overlap of the negative energy eigenvectors of $\mathbf{H}^-(U)$ with those of $\mathbf{H}^+(U)$. The inner product of the many body states is the determinant of this overlap matrix. In a similar fashion fermionic n-point functions are also expressed as determinants. Of course, in order to perform a useful numerical simulation in 4-D, efficient algorithms will be needed.

Formulation of the fermions using operators makes the role of topology transparent: One can rigorously show that $\mathbf{H}^-(U)$ always has an equal number of positive and negative eigenvalues (section 8 in [5]). This is not the case for $\mathbf{H}^+(U)$ which has, for non-perturbative gauge field configurations, different numbers of positive and negative eigenvalues. In such a situation the many body states have unequal fillings and the overlap vanishes. Insertion of an appropriate number of fermion creation or destruction operators inside the overlap can compensate for this mismatch and a non-zero expectation value for the operator inserted is obtained. By definition, such gauge fields carry a lattice topological charge equal to the number of inserted fermion operators counting the sign by fermion number in the auxiliary problem. This is how the overlap reproduces Property 2 of $\mathbf{C}(A)$. If one takes a lattice gauge field configuration with a non-zero topological charge and goes to finer lattices the matrices get bigger but the mismatch remains fixed corresponding to the topological charge of the approximated continuum gauge field configuration. The connection between this expectation value and the spatial distribution of the background gauge field originally discussed in [2] can be seen in complete detail in the overlap formalism. An example in 2-D illustrating this connection can be found in sections 10.4–10.6 in [5].

The mixed nature of the formalism is perfectly fine for all computations however the issue of unitarity and locality is not transparent. For this it is useful to recall a somewhat formal correspondence to a Grassmann path integral formulation. Another reason might be the familiarity with Feynman rules obtained usually in the path integral formalism. (Actually, Feynman rules are not necessary to do perturbation theory calculations and Schrödinger perturbation theory works quite well in practice [6,7].) It is possible to write down a path integral formalism for perturbative gauge fields (see section 6 in [5]). The fictitious time needed to project out the ground state is kept strictly infinite. This formalism cannot be extended to non-perturbative gauge fields (Refer to the comment in the last paragraph of section 6 in [5]). In fact a close variation of the path integral formalism in section 6 of [5] was erroneously "proven" to be a valid path integral representation of the overlap formula for all gauge fields in [8]. The error in [8] was anticipated in [5] and restated in [9].

As remarked before, the overlap formula is not gauge invariant in general on the lattice even in an anomaly free theory. Therefore one needs to integrate over the gauge orbits when performing the gauge integral on the lattice. It has been claimed in [8] that gauge averaging will make the overlap formula fail already in a 2-D U(1) theory restricted to the trivial orbit because this is the situation in which the waveguide model [10] failed and that the overlap formula has only been studied for smooth gauge fields. [11] also erroneously claims that the overlap formula has not been tested on any rough fields. Actually, the trivial orbit in 2-D U(1), for which the waveguide model failed, was considered in section 12 of [5]. There it is shown that the overlap and all naively gauge invariant n-point functions (allegedly corresponding to the "modified wave-guide" of [8]) are independent of the point on the trivial orbit at the lattice level. Therefore the overlap formalism has no problems for the fields claimed to destroy the waveguide model. In the waveguide model the chiral Dirac operator is represented by a finite square matrix and therefore inconsistent with property 2 which makes it a certain candidate for failure, one way or another.

The overlap formula has been successfully used in studying the Schwinger model in 2-D and the results are reported in this conference. The overlap formula is currently being used in a dynamical simulation of a 2-D chiral model. It would be necessary to develop clever algorithmic techniques to use the overlap formula in a 4-D computation.

Let us turn to property **1**. On the lattice, the issue of Lorenz invariance is not a severe requirement since it is broken anyhow. Still, associating an eigenvalue problem with $\mathbf{C}(A)$ is dangerous, since it might destroy Lorenz invariance also in the continuum. This is why we are happy that the overlap never associates an eigenvalue with the matrix representing $\mathbf{C}(A)$. Contrary to other approaches this matrix is generated dynamically, and its shape can be rectangular or square, depending on the gauge fields.

In the continuum a common practice for relating the chiral determinant to some eigenvalue problem is to add neutral spectator fermions of opposite chirality and make a Dirac operator out of $\mathbf{C}(A)$ and $\mathbf{C}^\dagger(0)$. Since $A=0$ is a Lorenz invariant but not gauge invariant configuration the new Dirac operator has Lorenz invariant, but not gauge invariant, eigenvalues [12]. This is fine for perturbative fields and can be used as a basis for defining the chiral determinant rigorously for continuum functions $A_\mu$. However, for topologically nontrivial connections this trick fails since the Dirac operator maps objects that have nontrivial sections in their left handed component and functions in their right handed component into objects with the opposite structure. So, this approach will not lead us to a non perturbative definition of 't Hooft vertex. While [12] does not discuss these vertices, the discussion in a follow up paper [13] makes no sense because of the above. Treating fermions in the continuum as above, while quite unlikely to become practical in any foreseeable future, is still incomplete even on the conceptual level when treating nontrivial topology. This seems to be ignored by most follow-ups of [12] listed in [14]. In addition to the above problem the continuum approach needs an interpolation of the lattice gauge fields. In [12] this seems to be done by first fixing the gauge on the lattice. An older proposal for interpolation is in [15] but the $U(1)$ example there contains discontinuities that are not removable by gauge transformations, in spite of what the text says. In general, a very local interpolation that maintains lattice gauge invariance, may run into difficulties by having to introduce many more discontinuities in the continuum connection than needed by the global topology and make the analysis of the continuum problem difficult. On the other hand a continuum approach that works in a fixed gauge could be useful; the non locality induced should be harmless in the anomaly free case once gauge invariance is restored in the continuum limit. We believe that the continuum approach can succeed (at least in principle and in the gauge fixed version) just because it avoids replacing the chiral operator by a fixed template matrix, or in equivalent words, because it uses an infinite number of fermions per lattice site [12].

R.N. was supported in part by the DOE under grant # DE-FG06-91ER40614 and grant # DE-FG06-90ER40561. H.N. was supported in part by the DOE under grant # DE-FG05-90ER40559.